\documentclass[a4paper,11pt]{article}
\usepackage{jcappub} 
\usepackage{lineno}

\usepackage{subfigure}
\usepackage{caption}
\usepackage{xcolor}
\usepackage{array}
\usepackage{graphicx}
\usepackage{tikz}
\usepackage{longtable}
\usepackage{booktabs}
\usepackage{url}

\title{\boldmath Direct reconstruction of the Reionization history from 21cm 2D Power Spectra}

\author[1]{Yannic Pietschke}
\author[1]{, Caroline Heneka}
\author[1]{, Tom Schlenker}
\author[1]{, Ayodele Ore}
\author[2]{, Benedikt Schosser}
\affiliation[1]{Institut für Theoretische Physik,
Universität Heidelberg, Philosophenweg 16, 69120 Heidelberg, Germany}
\affiliation[2]{Astronomisches Rechen-Institut, Zentrum für Astronomie der
Universität Heidelberg, \\
Philosophenweg 12, 69120 Heidelberg, Germany}

\emailAdd{pietschke@thphys.uni-heidelberg.de}

\abstract{The 21cm line from the spin-flip transition of neutral hydrogen (HI) provides a unique window into the Epoch of Reionization (EoR), the final phase transition of our Universe. 
The Square Kilometre Array (SKA) enables precise measurements of 21cm fluctuations that trace ionization, temperature, and density fluctuations of the intergalactic medium (IGM). Nevertheless, a direct reconstruction of the timeline of the EoR in terms of the progress of ionization remains an ongoing challenge due to the highly non-Gaussian nature and thus intractable likelihood of the 21cm signal. 
Here, we present \texttt{EoRFlow}, a simulation-based inference (SBI) framework for reconstructing the global neutral hydrogen fraction $x_{\mathrm{HI}}(z)$ directly from 2D cylindrically averaged power spectra (2DPS) of the 21cm signal. 
We validate our method on realistic mock datasets for SKA-Low. 
Bypassing the need for explicit likelihood formulations, our approach enables fast, unbiased posterior estimation of the $x_{\mathrm{HI}}$ evolution in narrow redshift slices, allowing for piecewise reconstruction of the global reionization history. By directly inferring the reionization history from 21cm power spectra, our framework provides a scalable and robust path forward for 21cm cosmology in the SKA era.
}

\begin{document}
\maketitle
\flushbottom

\section{Introduction}
Cosmic Dawn (CD), when the first luminous sources in the Universe form, and the Epoch of Reionization (EoR), when the first stars and galaxies ionize the surrounding intergalactic medium (IGM), represent key epochs in the history of the Universe, spanning redshifts from $\sim$5 to 20. The EoR in particular marks the Universe's last phase transition, from a neutral IGM to an ionized one. By mapping the forbidden spin-flip transition of neutral hydrogen (HI) that gives rise to the 21cm line, we can follow the IGM evolution and create 3D tomographic light-cones of the early Universe, thereby bridging the gap between the cosmic microwave background (CMB) and the late Universe uncovered by galaxy surveys. 

Radio interferometers such as the Precision Array for Probing the Epoch of Reionization (PAPER)~\citep{Parsons_2010}, and current experiments such as the Murchinson Widefield Array (MWA)~\citep{Tingay_2013}, the Hydrogen Epoch of Reionization Array (HERA)~\citep{DeBoer_2017} and the Low Frequency Array (LOFAR)~\citep{LOFAR} aim to detect the cosmic 21cm signal and have already provided upper limits on the 21cm power spectrum, see e.g.~\citep{Kolopanis_2019, Mertens_2020, Trott_2020, Yoshiura_2021, Abdurashidova_2022, Abdurashidova_2023, Ceccotti_2025}.
Furthermore, the Square Kilometre Array (SKA)\footnote{https://www.skao.int/en} is designed for the imaging of 21cm fluctuations and enables increased sensitivity for measuring power spectra in narrow redshift slices. This makes it possible to determine the timing and topology of the EoR, and thus the key parameters governing its evolution, in great detail.

A model-independent and therefore comparably direct way to probe the EoR is its evolution in terms of the fraction of neutral hydrogen $x_\mathrm{HI}$ as a function of redshift $z$. Observations of the Lyman-alpha forest place the end of the EoR at redshifts between $z \sim 5-6$~\citep{McGreer_2015, Eilers_2018, Yang_2020, Bosman_2022, Spina_2024}, and detect neutral IGM fractions of $x_\mathrm{HI} > 0.1$ at $z \geq 7$~\citep{Davies_2018, Greig_2022}.  However, the full reionization history remains uncertain, particularly whether reionization proceeded gradually and ended late or occurred more rapidly. 
Advancing our understanding of this process requires a direct measurement of the reionization history through the 21cm signal, not only to validate existing constraints but also to extend them to higher redshifts. 

The direct reconstruction of the EoR history $x_\mathrm{HI}(z)$ from the cosmic 21cm signal necessitates new analysis methods, due to the intractable likelihood, inherent non-Gaussianity, as well as foreground contamination and systematics of the signal. Even when using 21cm power spectra as (Gaussian) summary statistics to infer $x_\mathrm{HI}(z)$, the mapping between power spectra as a function of astrophysical and cosmological parameters to the neutral fraction remains non-trivial. Since fast simulators are available to generate the 21cm signal for varying astrophysics~\citep{Mes07,21cmfast11,Hassan:2016,Hutter:2021} 
and cosmology~\cite{Heneka2018, Liu2020}, we can use simulation-based inference (SBI) to relate fundamental parameters to observations~\cite{Brehmer:2019xox,Dax:2021tsq,Huertas-Company:2022wni,Villaescusa-Navarro:2022}. SBI is a flexible methodology and has been employed for the 21cm signal either with fixed summaries~\citep{Prelogovic_2024}, or with learned summaries for optimal and unbiased inference~\citep{Schosser_2025}. Recent progress in machine learning has even made it possible to learn new representations of the data that enable robust inference for transfer between different noise and model scenarios~\cite{Ore_2025}. 

A common statistic for analysis of the 21cm signal is the 2D cylindrically averaged power spectrum (2DPS). It goes beyond the standard 1DPS by separating spatial fluctuations across the sky, in modes $k_\perp$, from fluctuations along the line-of-sight, $k_\|$, and therefore improves upon parameter constraints compared to the 1DPS~\citep{Prelogovic_2024}. 
The 2DPS also allows for removal of foreground contaminations in the so-called foreground wedge~\citep{Liu_2014}. Most notably, it requires less integration time to achieve sensitivity for signal detection~\citep{Greig_2024}; a detection of the EoR 21cm signal with SKA via the 2DPS is thus expected to precede high-fidelity imaging.
We present \texttt{EoRFlow}, an SBI framework for reconstructing the global neutral hydrogen fraction $x_\mathrm{HI}(z)$ directly from a set of 21cm 2DPS. It is based on conditional normalizing flows and allows for an unbiased and fast inference of the joint posterior of the neutral hydrogen fraction for narrow redshift slices. Our method is able to robustly reconstruct the EoR history from simulated 21cm mock observations that include realistic SKA-Low noise. 

This paper is organized as follows: In Section~\ref{sec:data}, we provide an overview of the simulation data and the methods used for generating the 21 cm power spectra, as well as the generation of SKA mock observed light-cones. Section~\ref{sec:NN} details our SBI model, training and preprocessing of the data is outlined in Section~\ref{sec:train}. In Section~\ref{sec:Results}, we present the results for our reconstruction of the EoR history and an assessment of the SBI model performance. We finish with our summary and conclusions in Section~\ref{sec:Conclusion}. 

\section{21cm simulations and 2D power spectra} \label{sec:data}

For our network training database, we follow the procedure of~\citep{Neutsch_2022, Schosser_2025} and simulate 3D light-cones of the 21cm brightness temperature $\delta T_\mathrm{b}(\mathbf{x}, \nu)$, with on-sky coordinates $\mathbf{x}$ and frequency $\nu$,  using the publicly available semi-numerical code \texttt{21cmFAST v3}\footnote{https://github.com/21cmfast/21cmFAST}~\citep{21cmfast11, 21cmfast}. It generates initial conditions for the density and velocity fields in Lagrangian space and evolves them at first and approximate second order in perturbation theory based on the Zel'dovich approximation~\citep{zeldovich}. Ionized regions are identified with the 
excursion-set approach by using a
real-space top-hat filter at decreasing scales. A region is then considered as ionized if the fraction of collapsed matter $f_\mathrm{coll}$ exceeds the inverse ionizing efficiency $\zeta^{-1}$. In general the 21cm signal depends on the gas spin temperature $T_\mathrm{S}$, which is usually neglected in the so-called post-heating regime by assuming $\overline T_\mathrm{S} \gg T_\mathrm{CMB}$. We do not make this assumption here but instead evolve the spin temperature for all redshifts.

The resulting field of the 21cm brightness temperature depends on various cosmological and astrophysical parameters. Following~\citep{Neutsch_2022}, we vary six key parameters, describing cosmology, as well as the astrophysics of the epoch of reionization and cosmic dawn, in order to be able to train our network on a wide variety of reionization histories. These parameters are uniformly sampled and then passed into \texttt{21cmFAST} for light-cone simulation. A summary of the parameters with corresponding prior ranges can be found in Table \ref{tab:parameters}.

\begin{table}[h]
    \centering
    \begin{tabular}{l c}
        \toprule
        \textbf{Parameter} & \textbf{Prior Range} \\
        \midrule
        $\Omega_\mathrm{m}$ & $\mathcal{U}[0.2, 0.4]$ \\
        $m_\mathrm{WDM}$ & $\mathcal{U}[0.3, 10]$\\
        $\log_{10}{T_\mathrm{vir}}$ & $\mathcal{U}[4, {5.3}]$\\
        $\zeta$ & $\mathcal{U}[10,250]$\\
        $E_0$ & $\mathcal{U}[100,1500]$\\
        $\log_{10}{L_\mathrm{X}}$ & $\mathcal{U}[{38}, {42}]$\\
        \bottomrule
    \end{tabular}
    \caption{Summary of the simulation parameters along with their prior ranges. We consider the matter density $\Omega_\mathrm{M}$, the warm dark matter mass $m_\mathrm{WDM}$, the minimum virial temperature $T_\mathrm{vir}$, the ionization efficiency $\zeta$, the X-ray energy threshold for self-absorption by host galaxies $E_0$ and the specific X-ray luminosity $L_\mathrm{X}$. Note that $T_\mathrm{vir}$ and $L_\mathrm{X}$ are sampled log-uniformly while the other parameters are sampled uniformly.}
    \label{tab:parameters}
\end{table}

The simulations are generated with a box size of $200 \, \mathrm{Mpc}$ at a resolution of $1.43 \, \mathrm{Mpc}$. To exclude the most unrealistic reionization cases we require the resulting optical depth $\tau$ to be within $5\sigma$ of the Planck measurement $\tau = 0.054\pm 0.007$~\citep{Planck2018}. 
In total, we simulate 15000 light-cones.

In order to test our method on a more realistic dataset we transform the simulated light-cones into SKA mock observations using the public code \texttt{21cmSense}~\citep{21cmsense13, 21cmsense14}. Again following~\citep{Neutsch_2022, Schosser_2025}, the light-cones are split into coeval boxes at fixed redshift values. The thermal noise is then calculated for each box by sampling from a zero-mean Gaussian and added to the Fourier-transformed box. Finally, the full light-cone is reconstructed in real space. For thermal noise, we assume 1080h of integrated SKA-Low stage 1 observations. \texttt{21cmSense} provides three different foreground setting: optimistic, moderate and pessimistic. In this work, we will focus on two distinct scenarios. Firstly, we consider an optimistic scenario, in which we utilize the full design baseline AA4 for SKA-Low, that includes all 512 antennas, together with the optimistic foreground setting. In this case the 21cm foreground wedge is assumed to extend up to the primary field of view of the instrument. We will refer to this scenario as 'AA4 opt'.
Additionally, we explore a less optimistic scenario, in which we use the AA$^*$ telescope configuration for science commissioning, including 307 antennas, together with the moderate foreground setting. Here, the foreground wedge extends 0.1Mpc$^{-1}$ beyond the horizon limit. We will refer to this scenario as 'AA$^*$ mod'. These scenarios are intended to bracket the range of possible noise conditions, from a conservative foreground model with a near-future telescope configuration to a more optimistic foreground model with the full telescope configuration.

We aim to investigate how well we can constrain reionization from 21cm 2D cylindrically averaged power spectra (2DPS) alone. The (dimensional) 2DPS is defined as~\citep{Prelogovic_2024}
\begin{align}
    \Delta^2_\mathrm{2D}(k_\perp, k_\parallel)=\frac{k_\perp^2 k_\parallel}{4\pi^2 V}\left< | \delta  T_\mathrm{b}(\mathbf{k}) |^2 \right>_{k_\perp, k_\parallel},
\end{align}
where $k_\perp$ and $k_\|$ are the wavenumbers perpendicular and along the line of sight, respectively. $\delta T_\mathrm{b}$ is the differential brightness temperature offset of the 21cm signal and $V$ is the volume normalization.

\begin{figure}
    \centering
        \centering
        \includegraphics[scale=0.5]{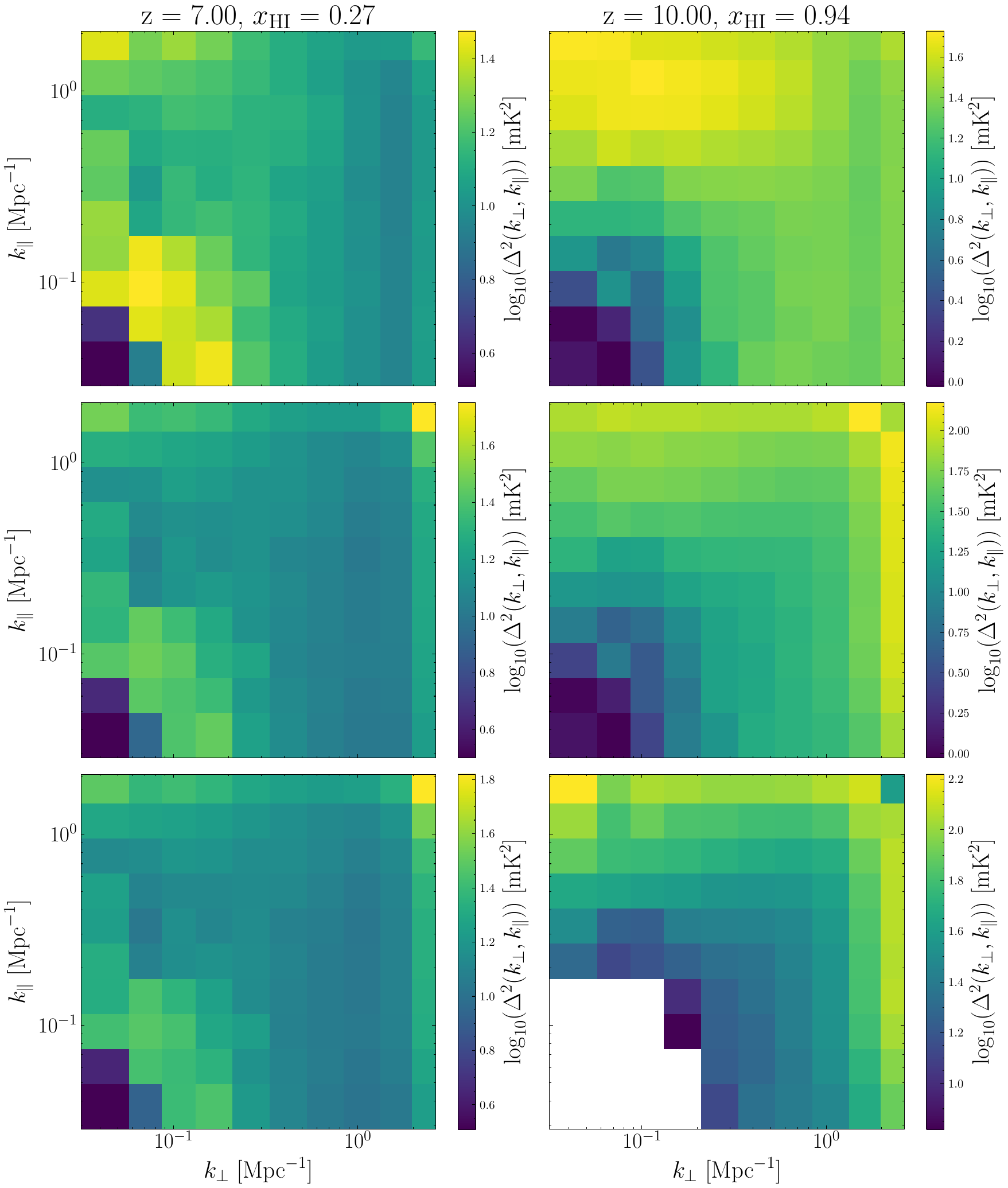}
    \caption{2DPS at redshifts $z=7$ (left) and $z=10$ (right) with corresponding neutral fractions $x_\mathrm{HI}$ for a fiducial light-cone with $\Omega_\mathrm{m}=0.31$, $m_\mathrm{WDM} =2 \, \mathrm{keV}$, $T_\mathrm{vir} =10^{5.2}$, $\zeta= 97.46$, $E_\mathrm{0} = 1143.21 \, \mathrm{eV}$ and $L_\mathrm{X} =10^{41.29} \, \mathrm{erg \, s^{-1}M^{-1}_\odot yr}$; shown are the 2DPS without noise (upper panels), for AA4 opt noise (center panels), and AA$^*$ mod noise (lower panels).}
    \label{fig: 2DPS}
\end{figure}

Motivated by current observational constraints of the Ly-$\alpha$ forest~\citep{Qin_2024} and by the James Webb Space Telescope (JWST)~\citep{Munoz_2024}, we here consider the redshift regime $z\in[5,12]$, thereby covering the full scope of reionization histories for a wide range of physically justified models. To capture the evolution of the neutral hydrogen fraction across this range, we compute the 2DPS for each light-cone at 15 redshifts spaced approximately $\Delta z \approx 0.5$ apart within that range. This choice balances resolution in redshift with signal-to-noise and computational tractability, enabling detailed reconstruction of $x_{\mathrm HI}(z)$ while maintaining feasible training times for our SBI framework.
For the 2DPS extraction we use the publicly available toolset \texttt{py21cmFAST-tools}\footnote{https://github.com/21cmfast/py21cmFAST-tools}. For each 2DPS we choose 10 bins in $k_\perp$ and $k_\|$, respectively, leading to a 3D input size per light-cone of $(15,10,10)$. The 2DPS at redshifts $z=7$ and $z=10$ for an example light-cone are shown in Figure~\ref{fig: 2DPS}. 

\begin{figure}
    \centering
    \includegraphics[scale=0.5]{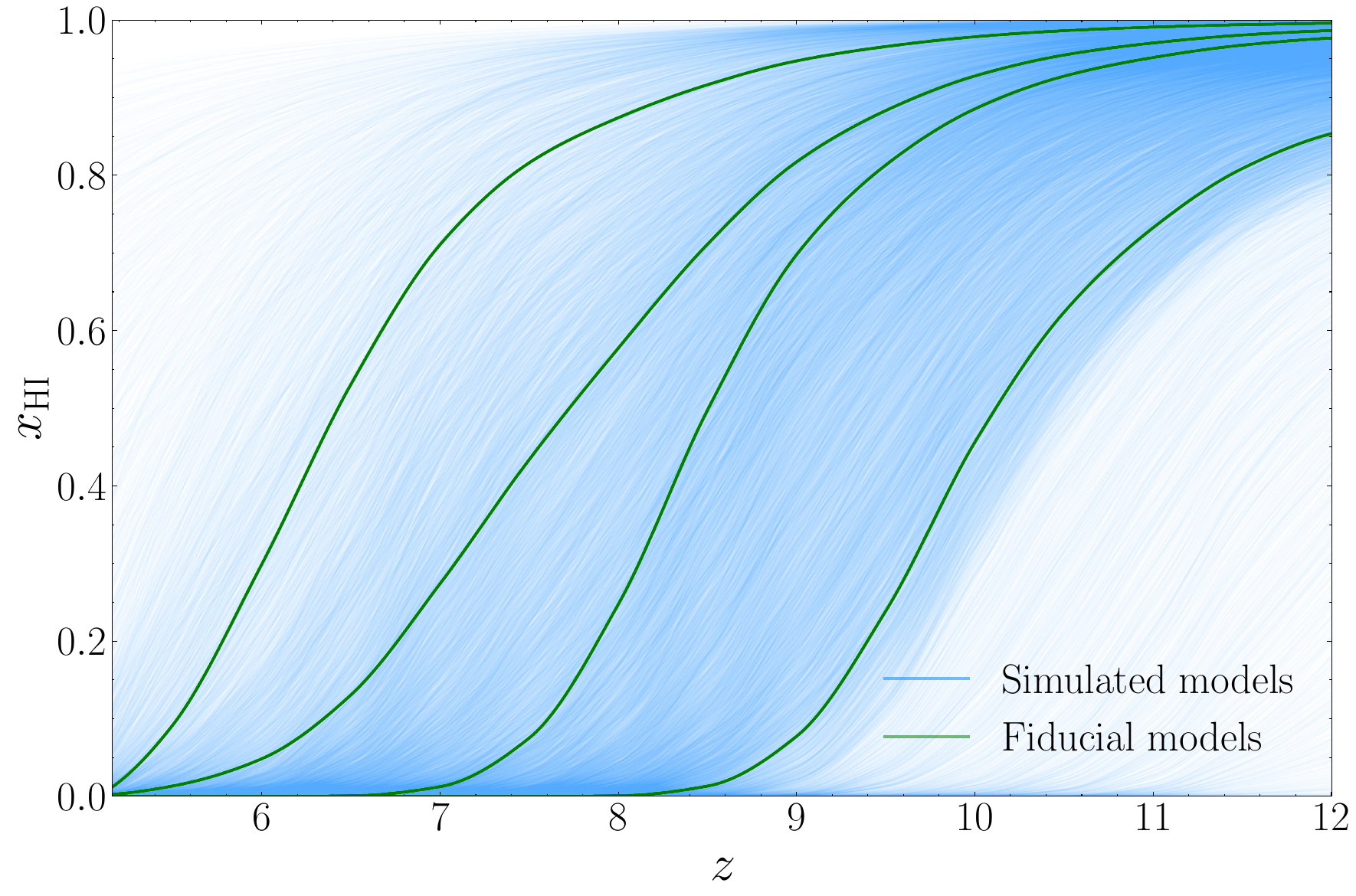}
    \caption{The 15000 simulated reionization histories of the dataset. The majority of models start and finish reionization entirely within the redshift regime $z\in[5,12]$, but some begin earlier or stop later. We retain these extreme models to ensure that our database remains as versatile as possible. The fiducial models used for analysis in Section~\ref{sec:Results} are highlighted in green (see Figure \ref{fig:inferred_timelines}).}
    \label{fig:dataset}
\end{figure}

As for our data labels, we compute the average global fraction of neutral hydrogen for each redshift $x_\mathrm{HI}(z)$. The reionization histories for our dataset are presented in Figure \ref{fig:dataset}.
It is important to note that sampling the simulation parameters in a uniform way leads to a non-uniform distribution of EoR histories. The majority of realizations reionize between $z=6$ and $z=8$, while only few models reach full ionization earlier or later. The non-uniform nature of the dataset provides a potential additional challenge for our inference framework; we note though (see Section~\ref{sec:calibration}) that our posteriors are well-calibrated across fiducial EoR histories.

\section{\texttt{EoRFlow} for reconstruction of the EoR history} \label{sec:NN}
\texttt{EoRFlow} infers the full posterior probability distribution of neutral hydrogen fractions $x_\mathrm{HI}(z)$, at different redshifts $z$, directly from the corresponding set of 2DPS. For that purpose, we employ a conditional invertible neural network (cINN) architecture. A similar method was used in~\cite{Schosser_2025} to infer cosmological and astrophysical parameters from simulated 21cm light-cones. In that case, a summary network based on a convolutional neural network (CNN) architecture, the 3D-21cmPIE-Net~\citep{Neutsch_2022}, was employed to reduce the data to a size that can be handled by the cINN. Here, we utilize the 2DPS as a summary statistic. We therefore omit the additional CNN embedding and instead directly use the respective 2DPS as a condition for the network. Note that one could also summarize the 2DPS further with the help of a CNN smaller than the 3D-21cmPIE-Net. However, we found that this does not improve the performance in terms of unbiased posterior estimation. 
We therefore use the flattened array of bins of the 2D power spectra as a condition for the inference network.

For the cINN, we use \texttt{FrEIA} (Framework for Easily Invertible Architectures)\footnote{https://github.com/vislearn/FrEIA}~\citep{Freia} to construct an invertible normalizing flow. The idea is to transform a complex parameter distribution, in our case of the neutral hydrogen fractions $x_\mathrm{HI}(z)$ of the training data, into a known distribution in latent space, typically a Gaussian. During training, the flow thus learns a mapping from the prior to the base distribution, under condition of the learned summary representation of the 2DPS. At evaluation the mapping is inverted to draw from the Gaussian latent space and output parameter values conditional on a given 2DPS, meaning it generates posterior parameter values. This is used to sample the full posterior distribution, allowing for robust and fast inference. 

To train the flow, the negative log-likelihood (NLL) loss function is minimized. It is given by
\begin{align}
    \mathcal{L_\mathrm{NLL}}=-\left< \frac{|\overline G_\phi (\theta|h(x))|^2}{2} + \log \left| \frac{\partial \overline G_\phi(\theta|h(x))}{\partial \theta} \right|\right>_{p(\theta,x)}
    \label{eq:NLL},
\end{align}
with data $x$ and parameters $\theta$. $h(x)$ represents the summary, that is the 2DPS of the corresponding 21cm data, and $G_\phi \, (\overline G_\phi)$ the flow (inverted flow). Samples from the joint distribution $p(\theta,x)$ are obtained by sampling the prior and then the likelihood as $\theta \sim p(\theta)$ and $x \sim p(x|\theta)$.
The first term in \eqref{eq:NLL} regularizes the network and the second term trains the Jacobian. 
A derivation of \eqref{eq:NLL} can be found in~\citep{Plehn_2024}. 

A fast evaluation of the Jacobian is ensured with a sequence of affine coupling layers~\citep{Dinh_2017}. 
In case a more expressive network is required, they can be interchanged for cubic~\citep{Durkan_Cubic} or rational quadratic splines~\citep{Durkan_Spline}. The exact architecture is shown in Appendix \ref{sec:B}.
For our task, we choose 10 affine coupling layers with 512 nodes each and ReLU activation. For hyperparameter optimization we used the \texttt{Optuna} framework\footnote{https://github.com/optuna/optuna}~\citep{optuna}.
All training and evaluation is performed in PyTorch~\citep{Pytorch} with the AdamW optimizer with default decoupled weight decay for regularisation~\citep{AdamW}.

\section{Training and preprocessing}\label{sec:train}

To ensure stable training and enhance inference performance, the 2DPS and corresponding redshifts per slice are min-max normalized. The redshift information helps the network contextualize the power spectra within the timeline of the EoR. The normalized redshifts are appended to the corresponding flattened 2DPS array before the set of 2DPS arrays at 15 different redshifts is concatenated. 
During training,  we apply to the labels $x_\mathrm{HI}$ a logit, or inverse sigmoid, transformation given by
\begin{align}
    \mathrm{logit}(x) = \ln{\frac{x}{1-x}},
\end{align}
where $x\in(0,1)$ is mapped to $\mathbb{R}$. 
During inference, we inversely apply the sigmoid transformation 
\begin{align}
    \sigma(x) = \frac{1}{1+\exp^{-x}},
\end{align}
to map the latent domain to the physical domain $x_\mathrm{HI}\in [0,1]$.

\begin{figure}[h!]
    \centering
    \includegraphics[width=\columnwidth]{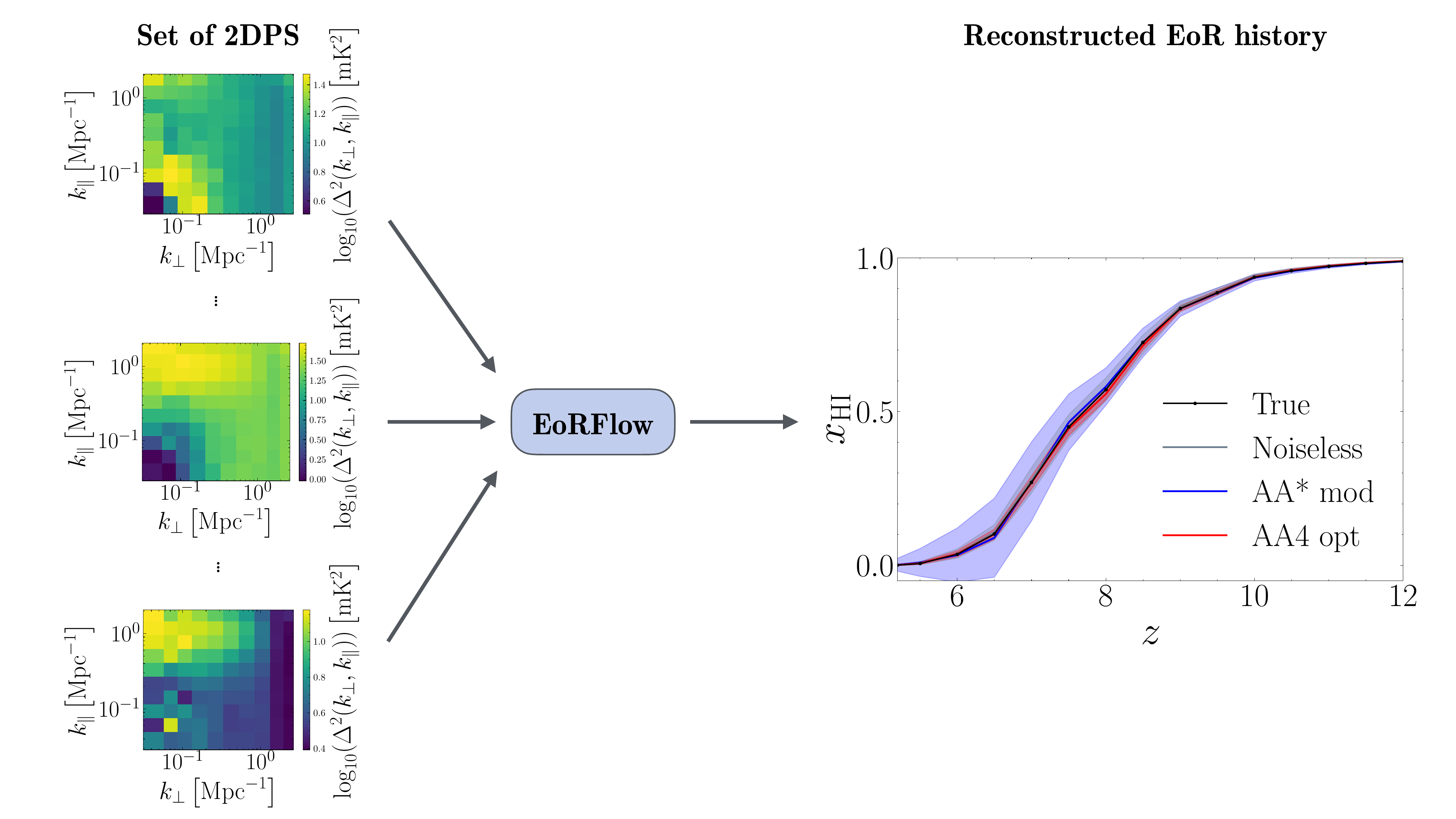}
    \caption{Illustration of the \texttt{EoRFlow} inference framework developed in this work. A set of 2DPS at 15 different redshift bins from $z=5.2$ to $z=12.0$ (see Section~\ref{sec:data} for a description of the data) is fed into \texttt{EoRFlow} for posterior estimation and reconstruction of the EoR history, see Section~\ref{sec:data} for a description of the 2DPS as well as noise and array configurations, AA$^\star$ mod and AA4 opt, and Section~\ref{sec:NN} for a detailed description of the \texttt{EoRFlow} model. }
    \label{fig:framework}
\end{figure}

For the database, we simulate 15000 light-cones and extract the 2DPS and $x_\mathrm{HI}(z)$ labels (see Section \ref{sec:data}). We use 10000 light-cones for training, 3000 for validation and 2000 for testing. Here, we perform two studies: a model trained on simulations without noise, and a model trained on the same simulations with noise added for a 1080h SKA-Low stage 1 observation, as described above. 
For the dataset without noise, we found that adding a small amount of Gaussian noise ($\sigma \sim 0.05$) to the power spectra stabilizes training and improves performance. The noise is added at random during training to the 2DPS, by sampling from a Gaussian $\mathcal{N}(0,\sigma^2)$.
The set of 15 2DPS, together with their redshift information, is passed to the flow as a condition.
An illustration of our inference framework is shown in Figure~\ref{fig:framework}. 

We train with a batch size of 16. An initial learning rate of 0.001 is used, which is halved every 10 epochs without improvement in the loss, as well as early stopping to avoid overfitting. The model for AA4 opt and AA$^*$ mod mocks is trained for 135 and 211 epochs, respectively. For noiseless simulations we train for 124 epochs. On one Nvidia A30 GPU training takes about 50min. The loss curves are shown in Appendix~\ref{sec:B}.

\section{Results} \label{sec:Results}
In order to reconstruct the EoR history, we train \texttt{EoRFlow} on 13000 21cm simulations, split into 15 2DPS computed at redshifts in the regime $z\in [5,12]$. This choice balances redshift resolution with signal-to-noise and computational tractability. For details on training and preprocessing methods see Section~\ref{sec:train}. Once trained, we evaluate our model on a test set of 2000 additional simulations. We split the evaluation of results into three parts. We start by inferring the 15-dimensional joint $x_\mathrm{HI}$ posteriors for test light-cones and show fiducial examples in Section~\ref{sec:posteriors}. To assess the calibration of those posteriors we compute the empirical coverage with simulation-based calibration in Section~\ref{sec:calibration}. Finally, in Section~\ref{sec:reconstruction}, we use the inferred posteriors to reconstruct the full EoR history.

\subsection{Inferring posteriors for the neutral fraction $x_\mathrm{HI}$} \label{sec:posteriors}
\begin{figure}[h!]
    \centering
    \includegraphics[width=\columnwidth]{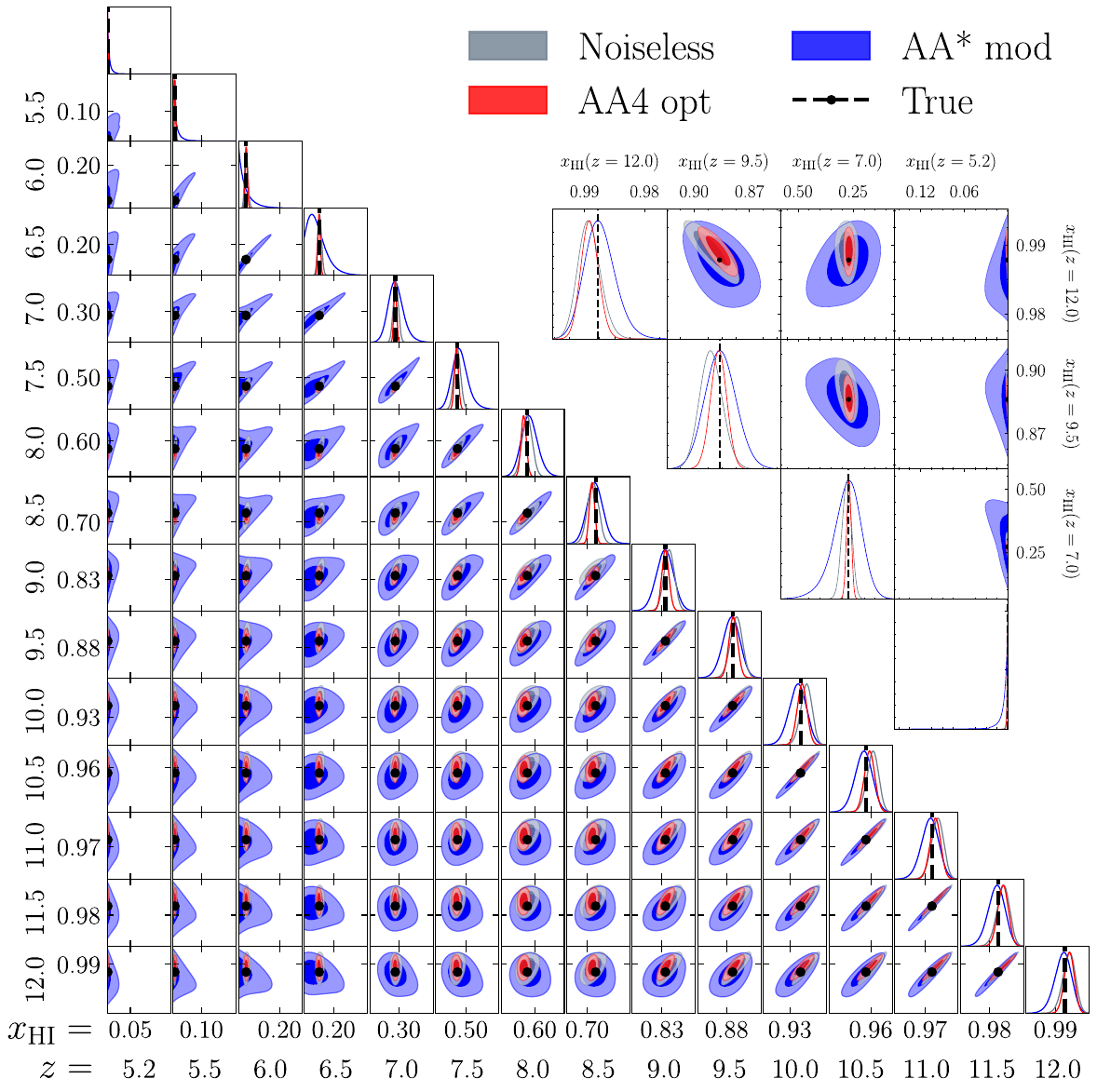}
    \caption{Marginalised posteriors for neutral fractions $x_\mathrm{HI}(z)$ for a randomly chosen set of parameters, $\Omega_\mathrm{m}=0.31$, $m_\mathrm{WDM} =2 \, \mathrm{keV}$, $T_\mathrm{vir} =10^{5.2}$, $\zeta= 97.46$, $E_\mathrm{0} = 1143.21 \, \mathrm{eV}$ and $L_\mathrm{X} =10^{41.29} \, \mathrm{erg \, s^{-1}M^{-1}_\odot yr}$. The shadings indicate 68$\%$ and 95$\%$ CI. Red shows the posterior derived from mock 2DPS including AA4 opt noise, blue corresponds to AA$^*$ mod noise and grey to the noiseless case for comparison; the black dot denotes the fiducial values. The zoom-in panel (upper right) shows for improved clarity a selection of contours with the same axis scaling as the scaling of the corresponding panel in the full triangle plot.
    }
    \label{fig:corner_main}
\end{figure}

To generate the full posterior for a fiducial EoR model as represented by a set of 2DPS, either for noiseless simulations or mock observations, we generate 1000 Gaussian samples and pass them through the inverted flow model, trained on either noiseless simulations or mocks, to obtain the posterior values of the corresponding neutral hydrogen fraction. This yields a 15-dimensional joint posterior for the full $x_\mathrm{HI}(z)$ distribution. The network evaluation for a single set of 2DPS takes about $0.5\,$s on one Nvidia A30 GPU; on a CPU (Intel i7-12700) evaluation takes about $5\,$s.
In Figure~\ref{fig:corner_main} we present the $x_\mathrm{HI}$ posteriors inferred from 2DPS calculated from a randomly drawn fiducial light-cone.  
The figure shows marginalized posterior distributions with 1- and 2$\sigma$ confidence regions, where the grey contours correspond to the model trained and evaluated on noiseless simulations. The red and blue contours represent the model trained and evaluated on AA4 opt and AA$^*$ mod mock simulations, respectively, bracketing different versions of realistic observational noise. The fiducial parameter values are shown as black dots. Note that in the main corner plot, the scale of the $x_\mathrm{HI}$ axes varies between redshift bins, as the confidence contours at lower redshifts are significantly tighter than at higher redshifts. To account for this, we include zoomed-in, to-scale examples in the upper right corner.

We find robust inference performance for both the noiseless and mock cases, meaning that the inferred posterior distributions consistently recover the true values within their respective uncertainty intervals. Specifically, the true neutral fraction values lie within the $2\sigma$ contours for the majority of tested scenarios, indicating that our model accurately estimates the uncertainty associated with each prediction (see also Section~\ref{sec:calibration} on calibration regarding both accuracy and precision, and the parameter scatter plot provided in Appendix~\ref{sec:A}). Moreover, we find no systematic bias in our predictions, i.e. there is no tendency toward over- or underestimating the neutral fraction.

Interestingly, the inferred posteriors for the AA4 opt mock data with noise included are typically slightly narrower compared to the noiseless case. This indicates that the presence of realistic observational noise can act as a regularizer during training, helping the network generalize better and thus yielding tighter constraints on the neutral hydrogen fraction. This counter-intuitive effect arises because noise prevents the network from overfitting to very subtle, simulation-specific features that would not be present in real observational data. For AA$^*$ mod noise the contours grow larger again. This is due to the less optimistic foreground avoidance scenario that removes highly foreground-contaminated modes, leading to information loss. For a more detailed discussion on the effect of noise on inference results and regularization see Appendix~\ref{sec:C}.
At the lowest redshifts, the posterior distributions correctly approach the physical boundary at $x_\mathrm{HI}=0$, clearly showing that \texttt{EoRFlow} respects the physical constraints of the neutral fraction. Looking at two neighboring redshift slices we see that the contours are tilting towards 45$^{\circ}$ illustrating a strong correlation, whereas the contours for widely separated slices appear uncorrelated. The model therefore correctly learns the correlations between the $x_\mathrm{HI}$-values at different redshifts.
More posterior examples as well as a plot of all predicted $x_\mathrm{HI}$-values can be found in Appendix~\ref{sec:A}.

\subsection{Calibration}\label{sec:calibration}
\begin{figure}[h!]
    \centering
    \includegraphics[scale=0.5]{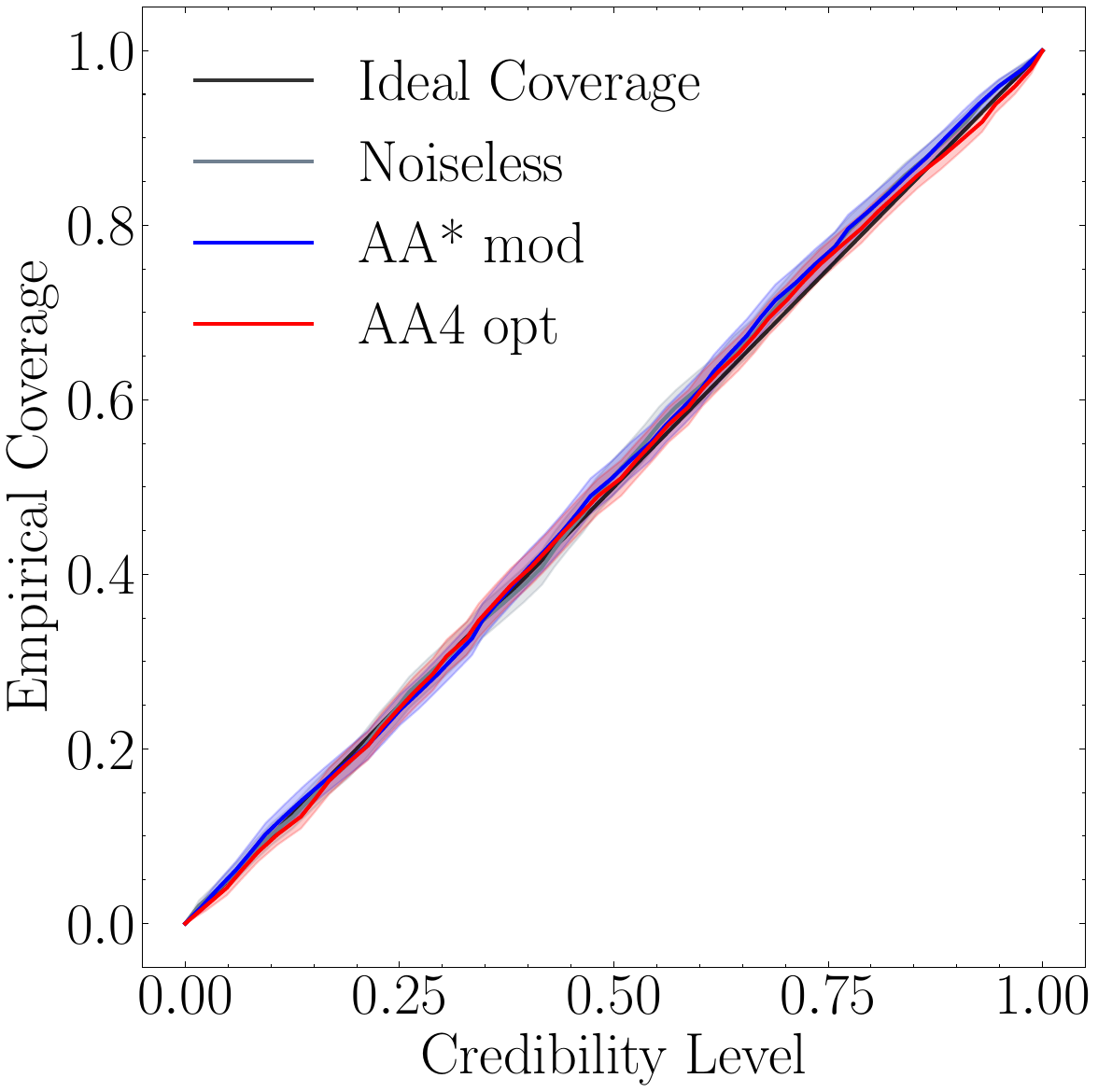}
    \caption{Empirical coverage of the posteriors for the model trained with AA4 opt noise (red), AA$^*$ mod noise (blue) and without noise (grey). Error bands are estimated based on a binomial distribution. The ideal coverage is represented by the black diagonal line.}
    \label{fig:coverage}
\end{figure}

For a more quantitative evaluation of the posteriors inferred with \texttt{EoRFlow}, we compute the empirical coverage for each credibility level using all test samples, which is shown in Figure~\ref{fig:coverage}. The coverage measures the frequency at which the true values fall within a given credibility interval, thereby quantifying how well the inferred posterior uncertainties reflect the true uncertainties. Ideal coverage, indicating perfect calibration, is represented by a diagonal line. If the empirical coverage lies above (below) this line, the inferred posterior is under-confident (over-confident), meaning that uncertainties are overestimated (underestimated). Error bands for the coverage can be obtained from a binomial distribution. Each sample is either within or outside of the $\alpha$-credible region. The expected variance due to the finite sample size $n$ can therefore be estimated by the binomial distribution $B(n,\alpha)$.

In our analysis, we find nearly ideal coverage for all models trained and evaluated on noiseless simulations, as well as for the ones trained and evaluated on mock observations with realistic noise. This demonstrates that our posteriors are very well calibrated, reliably reflecting the true uncertainties associated with each prediction. Additional confirmation of this calibration can be found in the scatter plots provided in Appendix~\ref{sec:A}, which visually show that the inferred values cluster closely around the true values, and that the size of the inferred uncertainty intervals (shown as error bars) consistently matches the actual deviations.

\subsection{Reconstructing the EoR history}\label{sec:reconstruction}
\begin{figure}[ht!]
    \centering
    \includegraphics[width=0.495\textwidth]{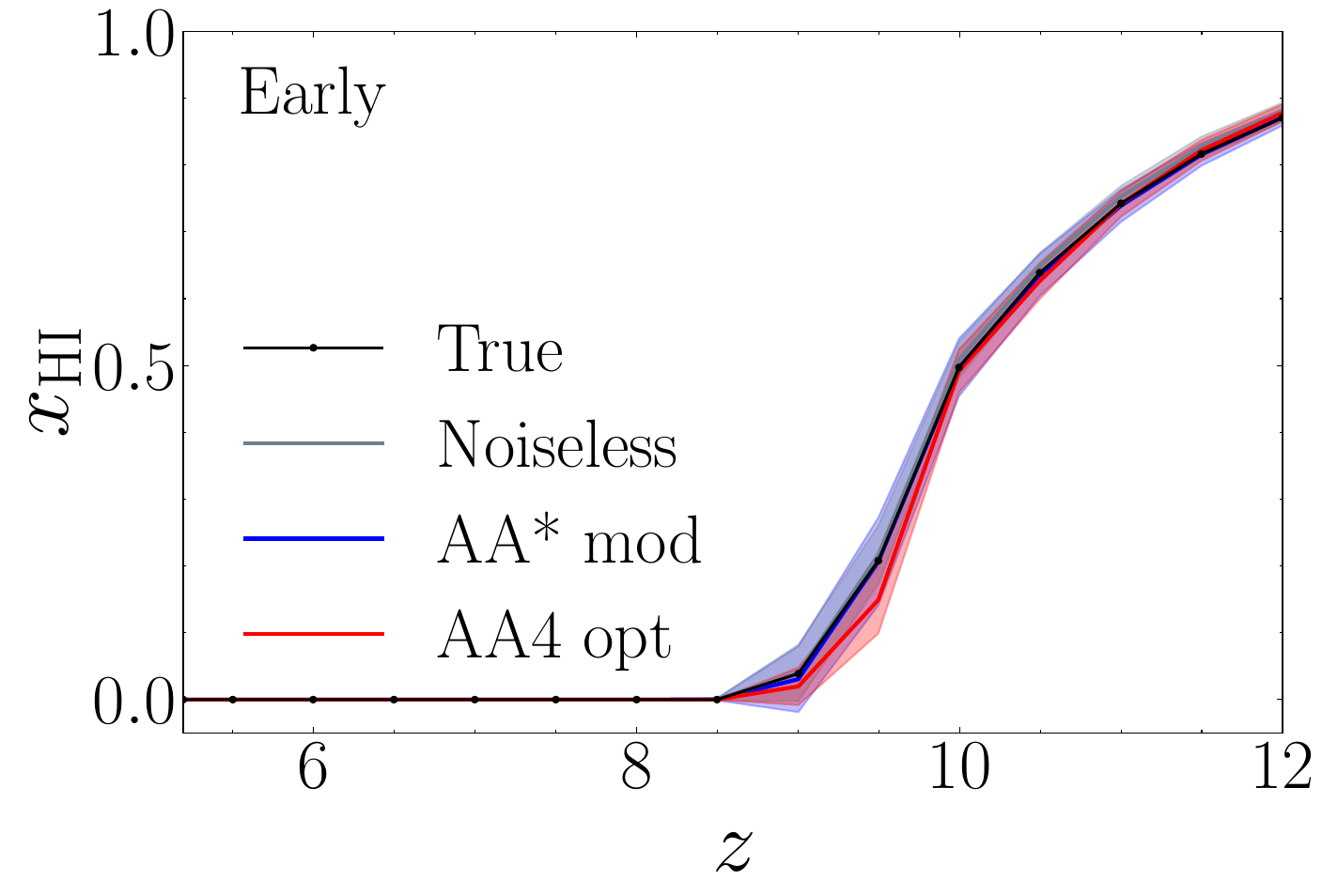}\hfill
    \includegraphics[width=0.495\textwidth]{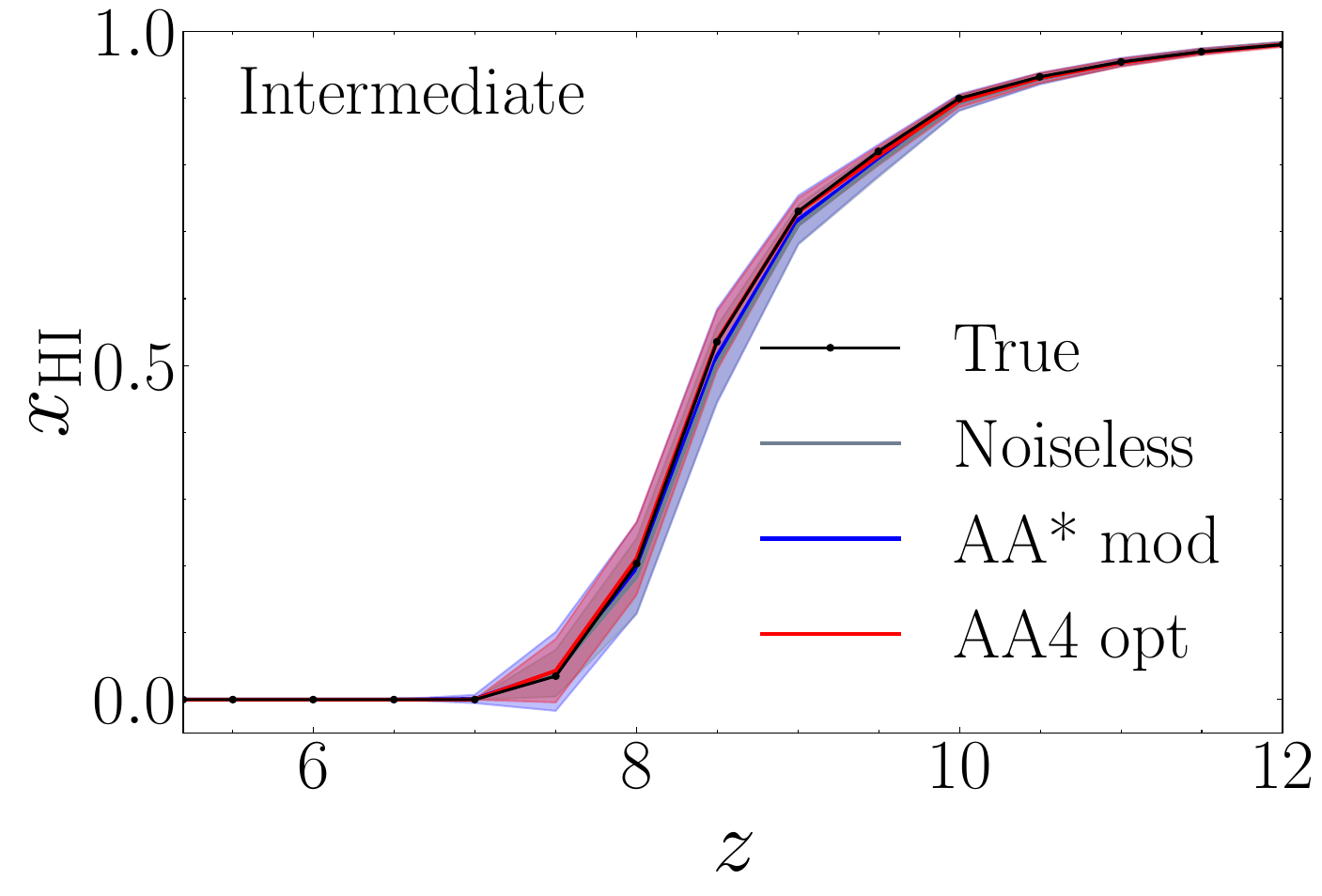}
    \\[\smallskipamount]
    \includegraphics[width=0.495\textwidth]{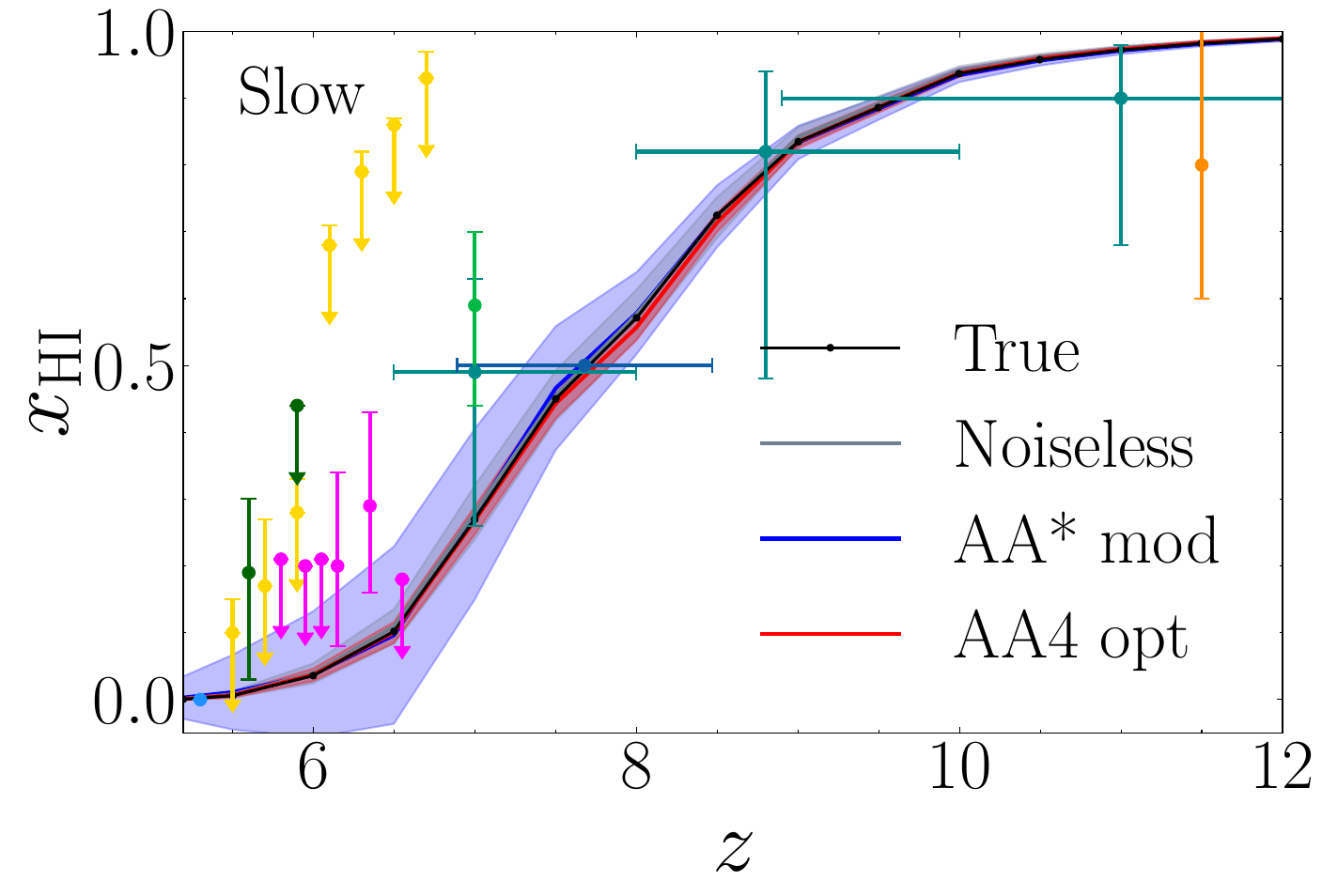}\hfill
    \includegraphics[width=0.495\textwidth]{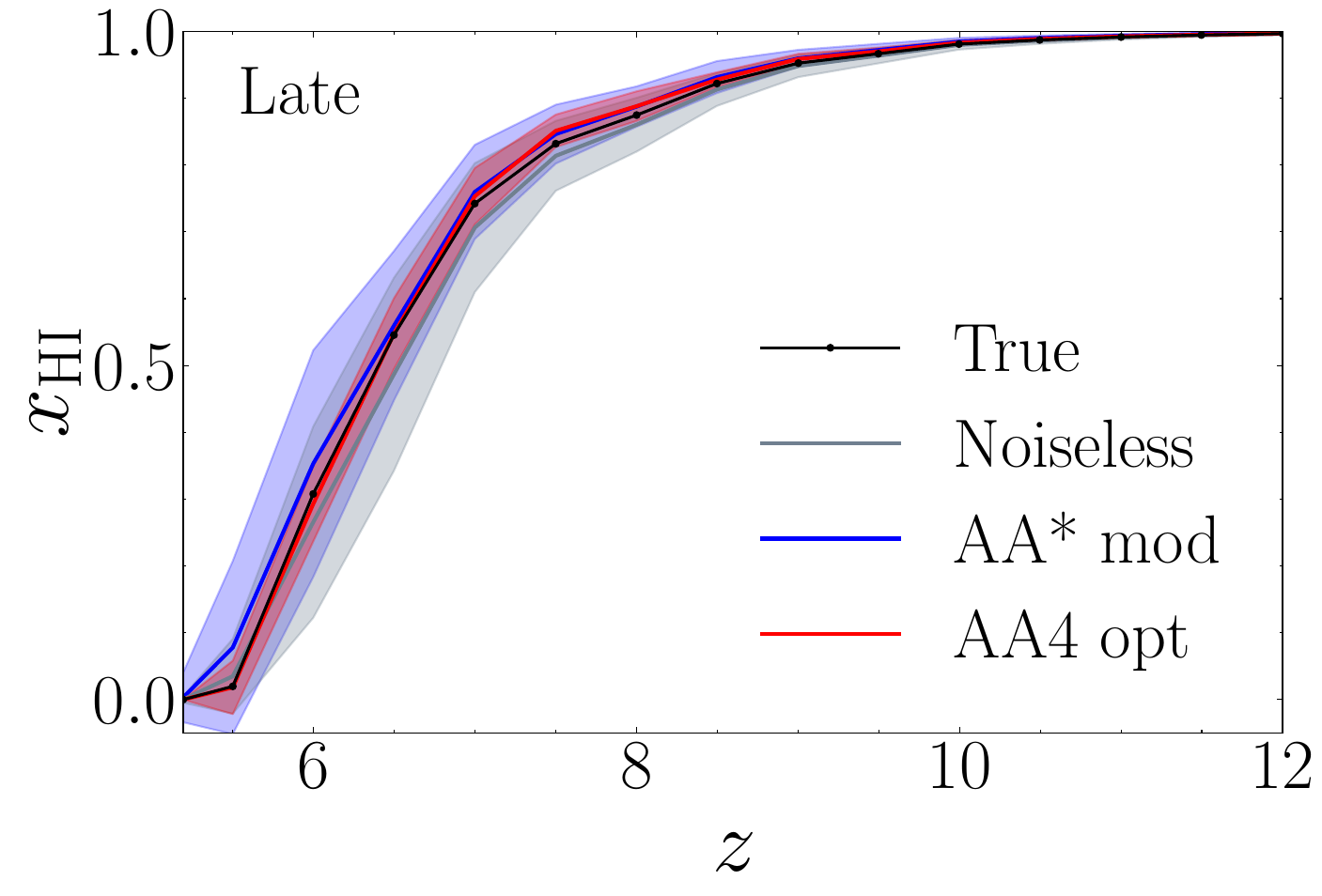}
    \caption{Reconstructed reionization histories for example test light-cones representative of early (upper left), intermediate/ slow (upper right and lower left), and late (lower right) reionization. The red, blue and grey lines show the models with AA4 opt, AA$^*$ mod noise configuration, and without noise, respectively. The shaded regions denote the $2\sigma$-errors. The simulation parameters for these models are: \textit{early:} $\Omega_\mathrm{m}=0.31$, $m_\mathrm{WDM} =9.85 \, \mathrm{keV}$, $T_\mathrm{vir} =10^{4.7}$, $\zeta= 155.93$, $E_\mathrm{0} = 1075.44 \, \mathrm{eV}$, $L_\mathrm{X} =10^{38.26} \, \mathrm{erg \, s^{-1}M^{-1}_\odot yr}$; \textit{intermediate:} $\Omega_\mathrm{m}=0.20$, $m_\mathrm{WDM} =4.93 \, \mathrm{keV}$, $T_\mathrm{vir} =10^{4.9}$, $\zeta= 209.47$, $E_\mathrm{0} = 186.30 \, \mathrm{eV}$, $L_\mathrm{X} =10^{41.57} \, \mathrm{erg \, s^{-1}M^{-1}_\odot yr}$; \textit{slow:} $\Omega_\mathrm{m}=0.31$, $m_\mathrm{WDM} =2 \, \mathrm{keV}$, $T_\mathrm{vir} =10^{5.2}$, $\zeta= 97.46$, $E_\mathrm{0} = 1143.21 \, \mathrm{eV}$, $L_\mathrm{X} =10^{41.29} \, \mathrm{erg \, s^{-1}M^{-1}_\odot yr}$ and \textit{late:} $\Omega_\mathrm{m}=0.32$, $m_\mathrm{WDM} =2 \, \mathrm{keV}$, $T_\mathrm{vir} =10^{5.5}$, $\zeta= 83.40$, $E_\mathrm{0} = 222.33 \, \mathrm{eV}$, $L_\mathrm{X} =10^{40.99} \, \mathrm{erg \, s^{-1}M^{-1}_\odot yr}$. For the slow reionization model we also show a variety of current Lyman-alpha constraints, as well as the latest Planck constraint of the reionization midpoint, represented by the colorful dots: \textit{gold:} \citep{Jin_2023}; \textit{light blue:} \citep{Bosman_2022}; \textit{dark cyan:} \citep{Tang_2024}; \textit{bright green:} \citep{Mason_2018}; \textit{dark green:} \citep{Spina_2024}; \textit{magenta:} \citep{Greig_2024_DW} and \textit{dark blue:} \citep{Planck2018}.
    }
    \label{fig:inferred_timelines} 
\end{figure}

To illustrate how we reconstruct the full EoR history from our inferred posterior, we take the 15-dimensional joint posterior obtained by \texttt{EoRFlow} and marginalize it to estimate the neutral fraction $x_\mathrm{HI}(z_i)$ individually at each redshift slice $z_i$. From each marginalized posterior, we derive the mean and quantify the uncertainty through its 2$\sigma$ interval. These estimates can then be compared directly to the true values, as visualized in the scatter plots presented in Appendix~\ref{sec:A} and by the fiducial EoR history in Figure~\ref{fig:inferred_timelines}.

Figure~\ref{fig:inferred_timelines} showcases the reconstructed neutral fraction as a function of redshift for four distinct EoR scenarios, ranging from late to early reionization. The inferred $x_\mathrm{HI}$ values at each redshift are plotted along with their corresponding 2$\sigma$ uncertainty intervals as shaded contours, with grey, red and blue lines again representing the noiseless simulations and the mock data with either AA4 opt or AA$^*$ mod noise, respectively. Fiducial $x_\mathrm{HI}$ values per bin $z_i$ are shown by black connected dots.

Overall, we find consistently accurate inference and reconstruction of EoR histories across the full range of tested reionization scenarios. Particularly noteworthy is the performance on less probable scenarios (see Figure~\ref{fig:dataset}), such as the late reionization example shown in the upper-left panel, where we observe appropriately larger uncertainty intervals. This demonstrates that \texttt{EoRFlow} successfully captures the increased uncertainty associated with EoR models that occur less frequently in our sample due to their sampling from astrophysical parameters, without any bias towards the more typical reionization histories in the training data.

Generally, we consistently find good performance across the entire test dataset regardless of how frequently or rarely specific reionization models appear in the training set, based on the sampling of models from astrophysical EoR parameters. The addition of realistic SKA-Low noise does not degrade the capabilities of our method. In fact, AA4 opt noise helps to regularize the network, leading to tighter, while still well-calibrated, constraints on the EoR history. Even when assuming the more conservative AA$^*$ mod noise scenario, \texttt{EoRFlow} is able to efficiently reconstruct the reionization timeline.

\section{Conclusion} \label{sec:Conclusion}

Current experiments like the HERA and LOFAR have already provided upper limits on the 21cm power spectrum. The SKA, in particular, enables increased sensitivity for measuring power spectra in narrow redshift slices and even image-based analysis of the 21cm signal. With these observations at hand, we can study the EoR history and evolution, as well as the parameters governing this crucial phase of our Universe in great detail.

However, a direct reconstruction of the reionization history remains an ongoing challenge. Due to the highly non-Gaussian nature and the resulting intractable likelihood of the 21cm signal, canonical inference methods that inherently assume an analytical likelihood function are difficult to apply.
SBI represents a promising alternative, leveraging advances in machine learning to extract information from simulated data without requiring any explicit likelihood assumptions.

In this work, we present \texttt{EoRFlow}, a novel SBI framework designed to reconstruct the global neutral hydrogen fraction $x_{\mathrm {HI}}(z)$, directly from the 2DPS of the cosmic 21cm signal. Our method employs conditional normalizing flows to efficiently bypass the need for explicit, analytically defined likelihoods, enabling robust and rapid posterior inference. By utilizing the 2DPS, we incorporate more morphological and directional information compared to conventional 1D power spectra, while maintaining computational efficiency and realism suitable for near-future observations.

We demonstrate the effectiveness of \texttt{EoRFlow} using realistic mock datasets representative of SKA-Low observations, covering a broad range of physically motivated reionization histories. Our method provides accurate and unbiased reconstructions of the neutral hydrogen fraction across the redshift range $z \in [5, 12]$, correctly capturing the uncertainty associated with both typical and less common reionization scenarios. Additionally, we show that the presence of realistic observational noise does not degrade the performance. In fact, for the AA4 opt scenario, the noise serves to regularize the network model, thereby improving stability and reliability for reconstruction of the EoR history. In the case of AA$^*$ mod noise, the posteriors grow larger due to the information loss caused by the assumption of a more extended foreground wedge. Despite this broadening, the constraints remain sufficiently tight to robustly reconstruct the reionization history from the 2DPS. 
The high accuracy and precision as shown by well-calibrated posterior coverage, and computational efficiency ($\sim 5\,$s per posterior evaluation) of \texttt{EoRFlow} make it particularly suited for processing the large datasets anticipated in the SKA era. It also offers flexibility, being adaptable to different summary statistics or even network-based learned representations. As we have shown previously, this is easily possible within our SBI approach also with 21cm lightcone (imaging) data both via fixed, as here with 2DPS, and learned summaries of the data~\citep{Neutsch_2022, Schosser_2025, Ore_2025}. By enabling direct inference of the reionization timeline from 21cm observations without restrictive assumptions, our approach represents a significant step forward in the analysis of the EoR and paves the way for fully exploiting the unprecedented sensitivity of SKA measurements. Constraining the EoR history from 21cm data is crucial for filling the gaps and extending the current constraints derived for example from Lyman-alpha forest observational probes, thus significantly enhancing our understanding of this critical epoch in cosmic evolution.

\acknowledgments

We would like to thank Daniela Breitman and Benedetta Spina for helpful discussions. We also thank the
anonymous referee for their insightful suggestions.
CH’s work is funded by the Volkswagen Foundation. This work was supported by the DFG under Germany’s Excellence Strategy EXC 2181/1 - 390900948 The Heidelberg STRUCTURES Excellence Cluster. 
The authors acknowledge support by the High Performance and Cloud Computing Group at the Zentrum für Datenverarbeitung of the University of Tübingen, the state of Baden-Württemberg through bwHPC
and the German Research Foundation (DFG) through grant no INST 37/935-1 FUGG.
BS's work is funded by Vector Stiftung.

\appendix

\section{Further posterior examples}\label{sec:A}
\begin{figure}[h!]
    \centering
    \includegraphics[width=\columnwidth]{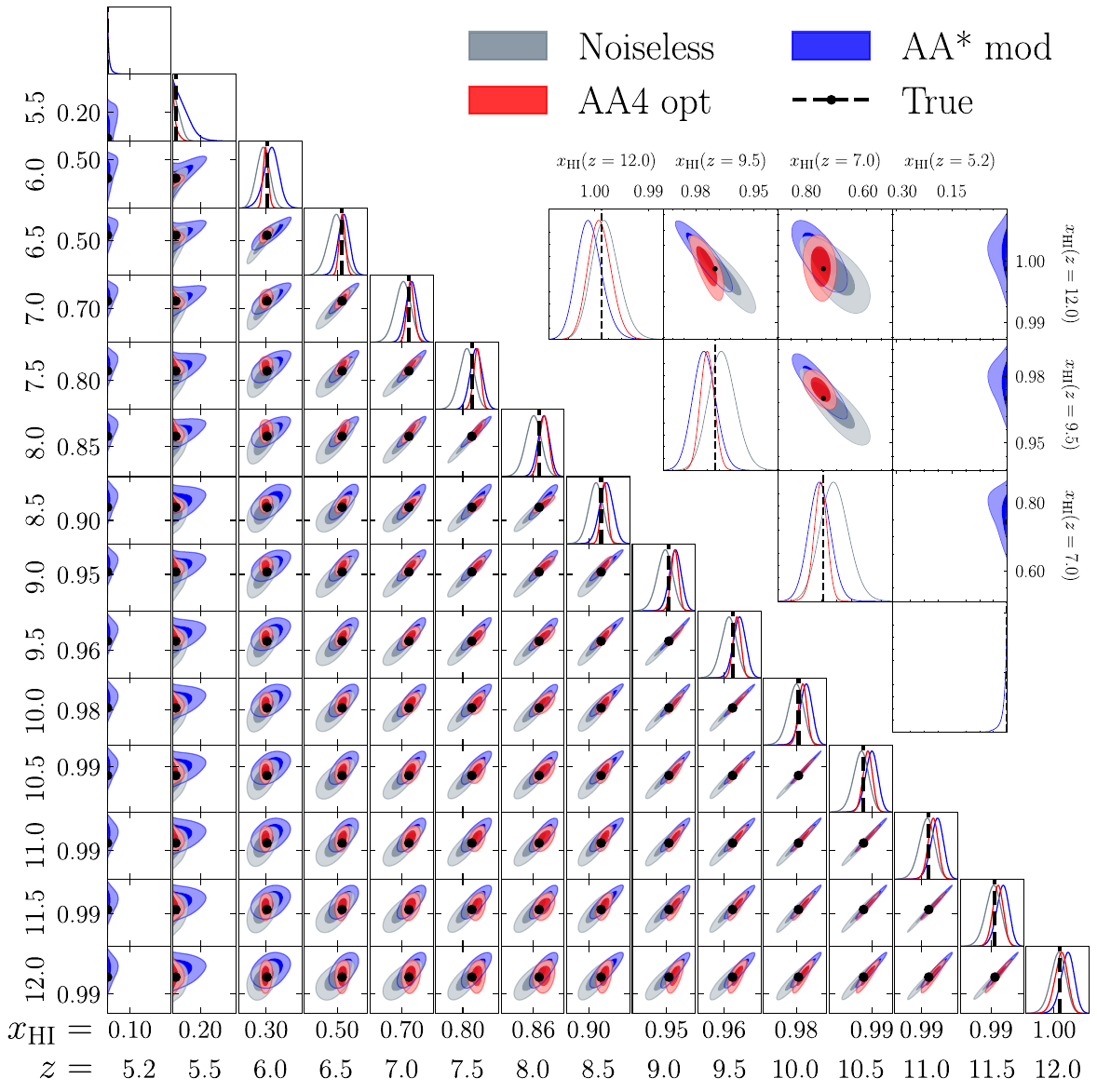}
    \caption{Marginalised posteriors for neutral fractions $x_\mathrm{HI}(z)$ for a randomly chosen set of parameters, $\Omega_\mathrm{m}=0.32$, $m_\mathrm{WDM} =2 \, \mathrm{keV}$, $T_\mathrm{vir} =10^{5.5}$, $\zeta= 83.40$, $E_\mathrm{0} = 222.33 \, \mathrm{eV}$ and $L_\mathrm{X} =10^{40.99} \, \mathrm{erg \, s^{-1}M^{-1}_\odot yr}$, yielding a late reionization model. The shadings indicate 68$\%$ and 95$\%$ CI. Red shows the posterior derived from mock 2DPS including AA4 opt noise, blue corresponds to AA$^*$ mod noise and grey to the noiseless case for comparison; the black dot denotes the fiducial values. The zoom-in panel (upper right) shows for improved clarity a selection of contours with the same axis scaling as the scaling of the corresponding panel in the full triangle plot.}
    \label{fig:corner_late}
\end{figure}

\begin{figure}[h!]
    \centering
    \includegraphics[width=\columnwidth]{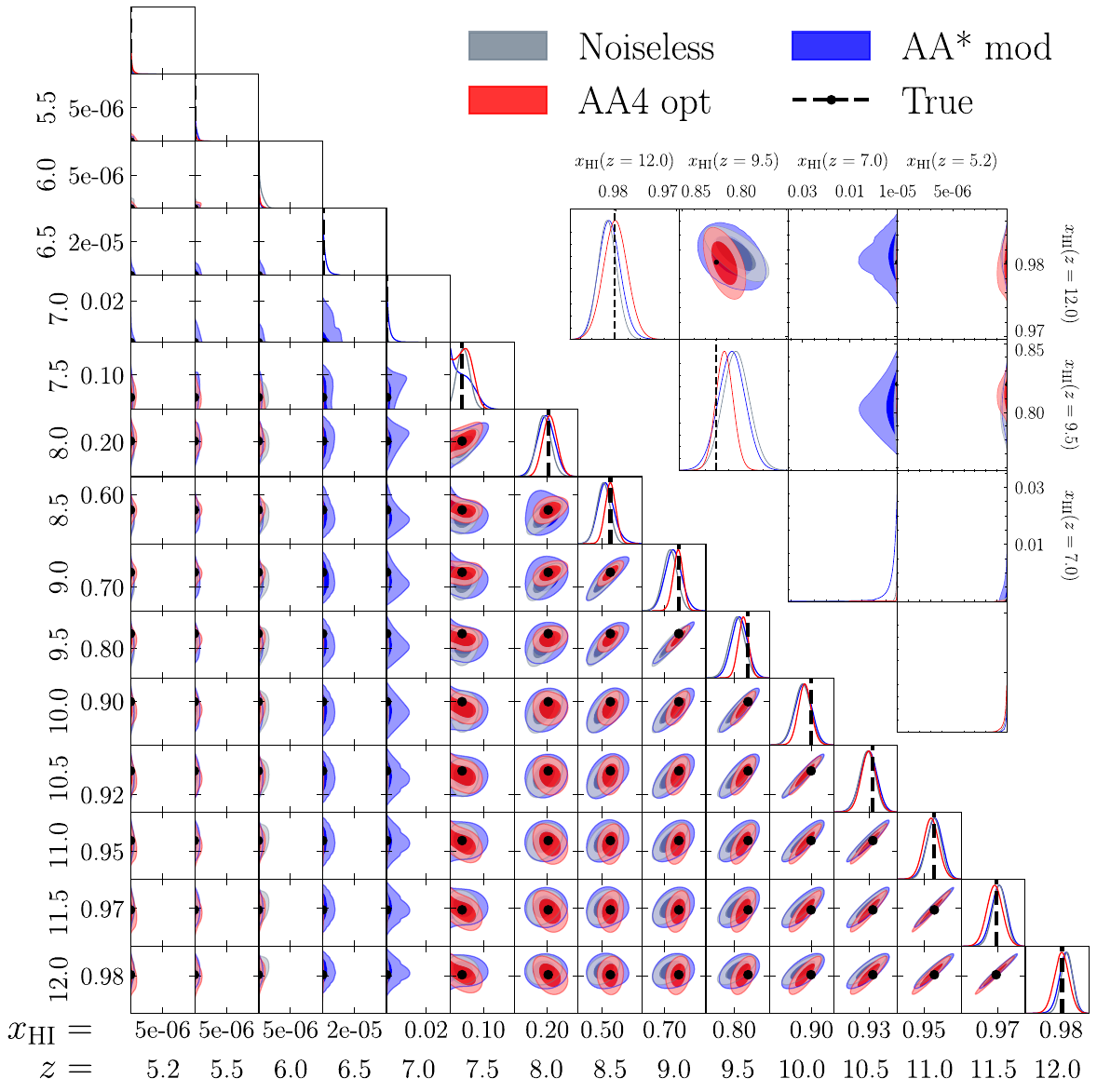}
    \caption{Marginalised posteriors for neutral fractions $x_\mathrm{HI}(z)$ for a randomly chosen set of parameters, $\Omega_\mathrm{m}=0.20$, $m_\mathrm{WDM} =4.93 \, \mathrm{keV}$, $T_\mathrm{vir} =10^{4.9}$, $\zeta= 209.47$, $E_\mathrm{0} = 186.30 \, \mathrm{eV}$ and $L_\mathrm{X} =10^{41.57} \, \mathrm{erg \, s^{-1}M^{-1}_\odot yr}$, yielding an intermediate reionization model. The shadings indicate 68$\%$ and 95$\%$ CI. Red shows the posterior derived from mock 2DPS including AA4 opt noise, blue corresponds to AA$^*$ mod noise and grey to the noiseless case for comparison; the black dot denotes the fiducial values. The zoom-in panel (upper right) shows for improved clarity a selection of contours with the same axis scaling as the scaling of the corresponding panel in the full triangle plot.}
    \label{fig:corner_mid}
\end{figure}

\begin{figure}[!ht]
  \centering
  \includegraphics[width=0.79\columnwidth]{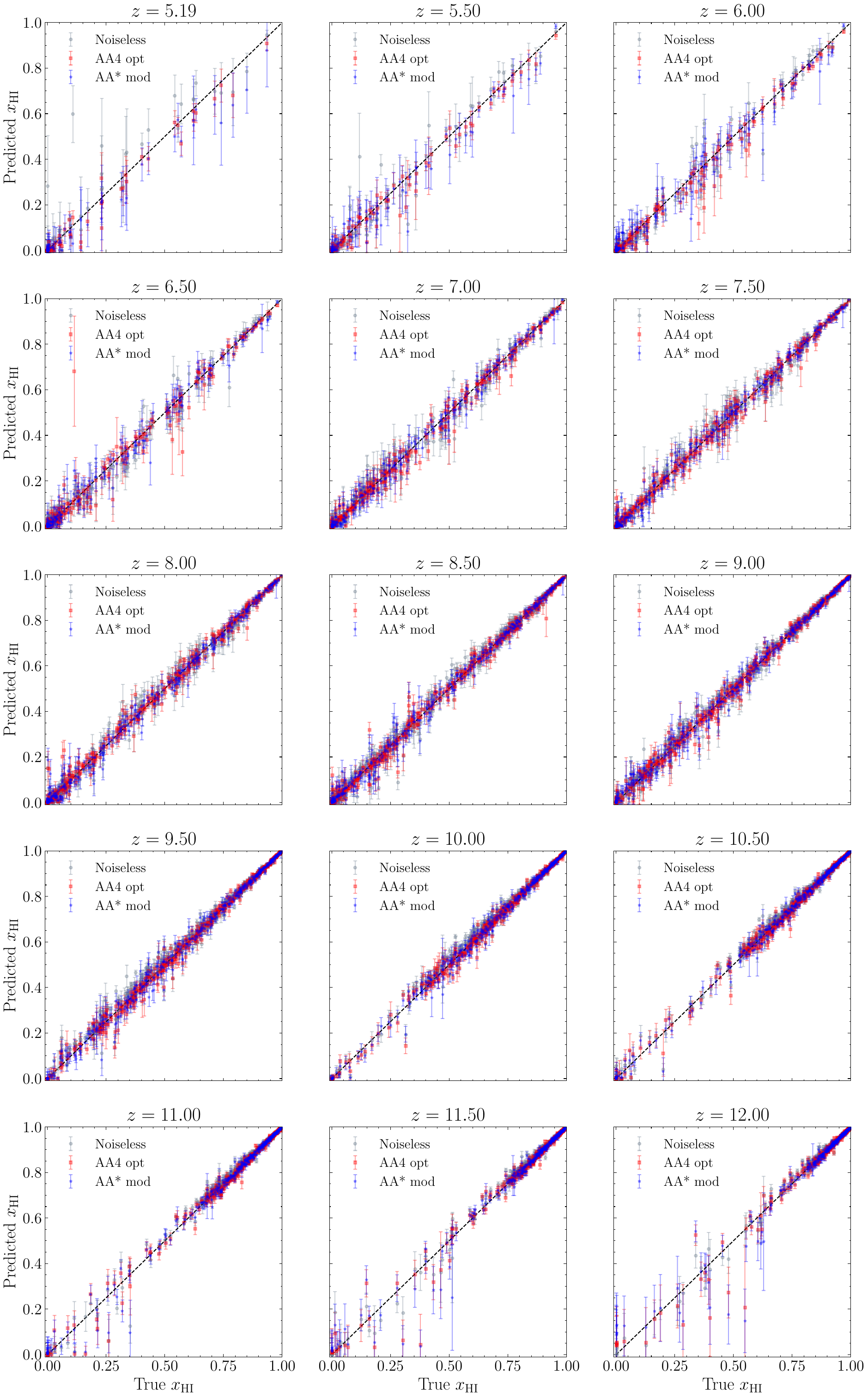}
  \caption{Scatter plot of all predicted samples including 1\,$\sigma$‐errors for 500 test light‐cones. The models trained with AA4 opt, AA$^*$ mod and without noise are shown in red, blue and grey, respectively. At low redshifts, almost all models have $x_\mathrm{HI}\approx 0$; higher values are rare and  therefore predicted, as wanted, with larger error bars. At high redshifts the inverse effect is seen. 
  }
  \label{fig:scatter}
\end{figure}

In this section we present further corner plots of the fiducial models from Figure~\ref{fig:inferred_timelines}. The corresponding plots are shown in Figures~\ref{fig:corner_late} and \ref{fig:corner_mid}. Furthermore, a scatter plot of all predicted samples of our model can be seen in Figure~\ref{fig:scatter}. 

\clearpage

\section{Network architecture and training loss}\label{sec:B}
\begin{figure}[h!]
    \includegraphics[width=0.55\columnwidth]{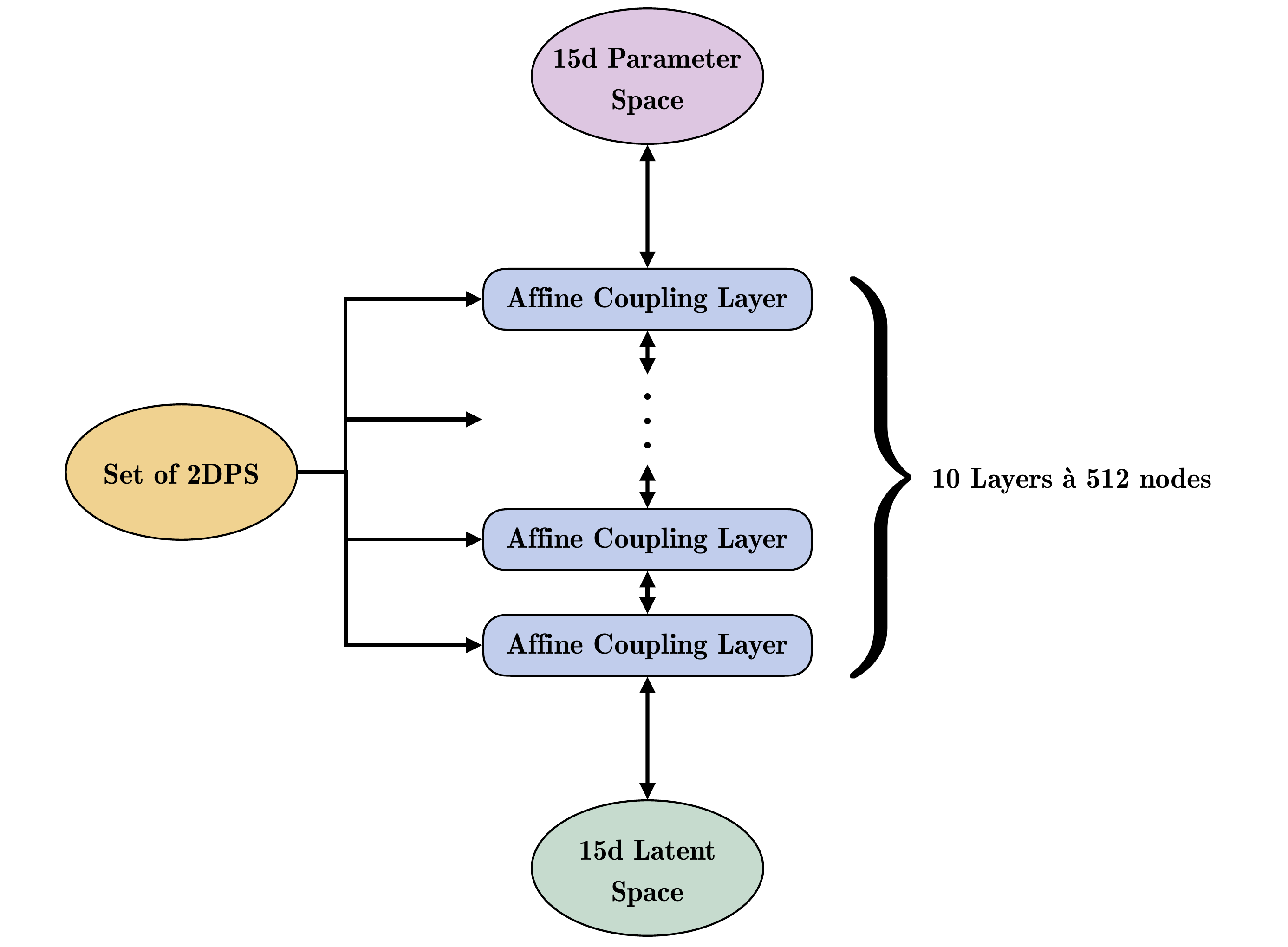}
     \includegraphics[width=0.43\columnwidth]{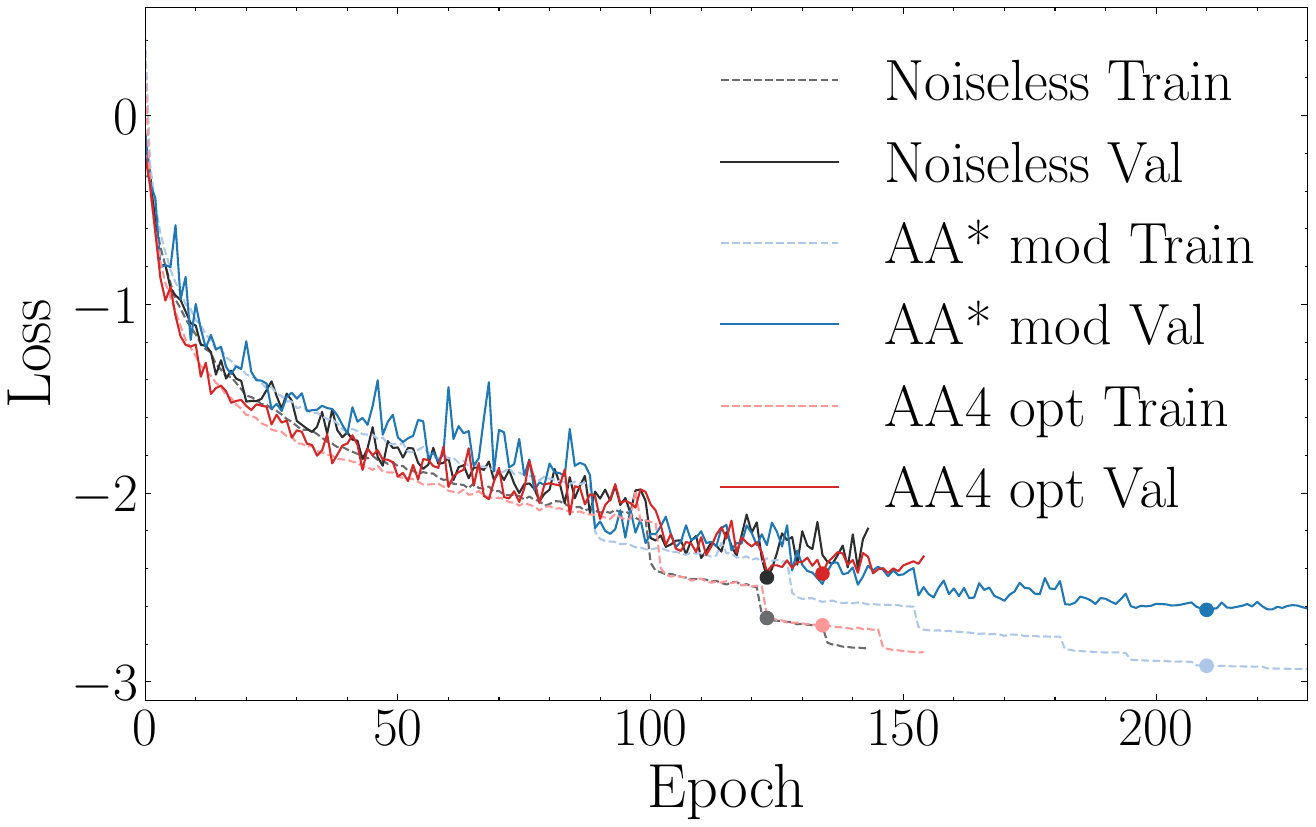}
    \caption{Left: Sketch of the \texttt{EoRFlow} architecture based on the \texttt{FrEIA}~\citep{Freia} package. We find 10 affine coupling layers with 512 nodes and ReLU activation to perform best for our application. For a more expressive network, if required, say in the case of a significantly larger number of redshift bins, affine coupling layers can be replaced by cubic or rational quadratic splines. Right: Loss curves for our models trained on mock and noiseless simulations, respectively. The training loss (train) curves are shown as dashed lines, while the solid lines represent the validation loss (val) curves. We utilize early stopping with a patience of 20 epochs. The best performing model trained on AA4 opt mock simulations is reached after 135 epochs, for AA$^*$ mod mocks after 211 epochs. For training without noise we find the best performing model after 124 epochs. These models are highlighted by the points in the corresponding color. Even though there is slight overfitting, these models generalize the best due to the minimum in validations losses which is also visible in their inference performance (see Section~\ref{sec:Results}).}
    \label{fig:architecture}
\end{figure}

In this section we present an overview of the network architecture of \texttt{EoRFlow}, as well as the loss curves of our training, in Figure~\ref{fig:architecture}. 

\section{Impact of noise on network training and inference results}\label{sec:C}
In this section, we briefly examine the effect of noise on the inference performance of our network. As discussed in Section~\ref{sec:posteriors}, we found that the posteriors obtained from a model trained and evaluated on noiseless simulations (including only a small Gaussian noise component with $\sigma \sim 0.05$ to stabilize training) were broader than those from a model trained on simulations with SKA AA4 noise and the optimistic foreground scenario. We interpret this as a regularization effect introduced by the noise. 
To test this hypothesis, we retrained the model on noiseless simulations, adding varying levels of Gaussian noise during training. We found that it is indeed possible to reproduce the results obtained with optimistic noise by adding an appropriate Gaussian noise level. Theoretically, this matching should occur, when thermal noise dominates over systematics for the optimistic case, at roughly $\sigma = 1$, which corresponds to the thermal noise component used in \texttt{21cmSense}, and we confirmed this empirically.
We further trained the model using the moderate noise configuration expected for the SKA AA$^\ast$, which incorporates reduced sensitivity (307 antennas instead of 512) and the, less optimistic, moderate foreground scenario. This setup represents a realistic scenario for the first SKA science data. In this case, we again achieved effective inference of the neutral hydrogen fraction across all redshifts, though the posteriors were broader than those obtained with the AA4 opt noise configuration.
To understand these results, it is useful to distinguish between two noise contributions.
Thermal (Gaussian) noise up to moderate levels regularizes the network by stabilizing training, smoothing the data and thus improving generalization. However, if the Gaussian noise level becomes increasingly large,  the network struggles to efficiently extract the 21cm signal from the noisy power spectra.
Systematic noise effects due to foregrounds and the removal of foreground-contaminated modes in the foreground wedge increase the derived uncertainties for increased wedge removal. In the AA$^*$ moderate foreground scenario, the removal of a larger wedge in k-space as compared to the optimistic scenario leads to broader posteriors, reflecting the loss of information.
Despite the larger uncertainties, the constraints remain sufficiently tight to robustly reconstruct the reionization history from the 2DPS.

\section*{Code Availability}
The code used for this paper can be found at \url{https://github.com/astro-ML/EoRFlow}.

\bibliographystyle{JHEP}
\bibliography{biblio.bib}

\end{document}